\documentclass[twocolumn,prl,a4paper,superscriptaddress]{revtex4}
\usepackage{comment}
\usepackage{color}
\usepackage{amsmath}
\usepackage{amssymb}
\usepackage{graphicx}
\usepackage{braket} 
\usepackage{esint}
\usepackage{adjustbox} 
\usepackage[utf8]{inputenc}

\usepackage[unicode=true,pdfusetitle,
bookmarks=true,bookmarksnumbered=false,bookmarksopen=false,
breaklinks=true,pdfborder={0 0 1},backref=false,colorlinks=true]
{hyperref}
\hypersetup{
	linkcolor=red,urlcolor=blue,citecolor=blue,anchorcolor=blue} 
\usepackage{ulem} 

\usepackage{dcolumn}\usepackage{bm}

\makeatother
\begin{document}
	
	\title{Thermalization of Isolated $BEC$ Under a $PT$-Symmetric Environment}
	
	\author{Javed Akram}
	
	\email{javedakram@daad-alumni.de}
	
	\affiliation{Department of Physics, COMSATS University Islamabad, Islamabad Pakistan}
	
	\author{Asad Hussain }
	
	\author{Muhammad Nouman}

	\affiliation{Department of Physics, COMSATS University Islamabad, Islamabad Pakistan}
	
		\author{Jameel Hussian}

	\affiliation{Department of Electronics, Quaid-i-Azam University, Islamabad, Pakistan}

	\date{\today}

\begin{abstract}
The postulates of the eigenstate thermalization hypothesis ($ETH$) expresses that the thermalization occurs due to the individual eigenstate of the system's Hamiltonian. But the $ETH$ put no light on the dynamics that lead  toward the thermalization. In this paper, we observe the thermalization of a Bose-Einstein Condensate ($BEC$) confined in an optical lattice potential that is embedded on the harmonic trap. Such optical lattice potential offers   local friction to the oscillating $BEC$. The spread in the temporal density plot of $BEC$ shows the thermalization of the $BEC$. Moreover, we observe that the presence of a $PT$-symmetric potential greatly influences the $BEC$ dynamics and the thermalization of the system. The presence of a $PT$-symmetric potential offers a way to manipulate the mean position of the $BEC$ to a desire location and for a desired length of time. 
	\end{abstract} 
	
	
	\maketitle
 
\maketitle

\section{Introduction} \label{sec1}
Confirmation of long-standing diverse ideas of condensed matter physics begins with the first realization of  Bose-Einstein condensates
 ($BEC$s) of dilute atomic gases \cite{Anderson-1995,PhysRevLett.75.3969,PhysRevLett.78.985,PhysRevLett.81.3811}. Among those ideas included the nature of superfluidity, the critical velocity for the beginning of the dissipation \cite{PhysRevLett.83.2502,PhysRevLett.85.2228}, quantization of vertices \cite{PhysRevLett.84.806,PhysRevLett.87.210403,PhysRevLett.88.010405}, the generation and dynamics of soliton waves  \cite{PhysRevA.93.023606,PhysRevA.93.033610,Akram_2016,Hussain2019}, and the impact of impurities for different practical applications \cite{PhysRevA.93.023606,Akram_2018,PhysRevA.93.033610}. 
Recently it also leads to probe the long-standing question of thermalization of an isolated quantum system both theoretically and experimentally \cite{Posazhennikova-2017,RevModPhys.83.863,Trotzky-2012}. In these studies,  the thermalization observed in double-well potential confinement under the influence of Josephson interaction \cite{Posazhennikova-2017}, and it was also  investigated experimentally in an optical lattice environment \cite{Trotzky-2012}.
\par In this paper, we study the thermalization of BEC in harmonic trap embeds  with optical lattice potential. We also investigate the impact of a $PT$-symmetric periodic potential on the thermalization of the BEC. We initially trapped BEC in a harmonic trap after achieving the equilibrium, we shift the harmonic potential minima for time $t>0$. At the same time, we switched on the optical lattice which is embedded in the harmonic trap.  Such a lattice potential offers  friction for the dipole oscillations of the $BEC$.  The idea of $PT$-symmetric potential represents the scenario in which there are alternative gain and lost regions in an optical lattice, e.g, as explained for  double-well confinement \cite{Hussain2020}. In an optical lattice, the transmission between adjacent wells is controlled by controlling the tunneling between the wells. The periodic  lattice potential here acts as a medium that absorbs the partial kinetic energy and partial potential energy of the $BEC$ entropy. 
That medium helps in bringing the $BEC$ into  thermal equilibrium. Furthermore, we also test the localization and thermalization of $BEC$  under a periodic $PT$-symmetric environment.
 The idea of non-Hermitian Hamiltonian obeying $PT$-symmetry was introduced by Bender and Boettcher \cite{PhysRevLett.80.5243} and this appears as an extension to quantum mechanics from a real to the complex domain. The $PT$-symmetric conditions are more physical than the earlier strict mathematical condition of hermiticity of the Hamiltonian for real eigenvalues. The operator "$P$" and "$T$" represent  parity reflection and time reversal, respectively. The operator $P$ acts on position and momentum operator as  $P$: $x \rightarrow-x$, $p \rightarrow -p$ and the time operator $T$ acts on position and momentum operator as $T$: $x \rightarrow x$, $p \rightarrow -p$, and $i \rightarrow -i$.
The remaining part of this paper is organized as follows. In Sec. \ref{sec2}, we describe the working models, like analytical, numerical, and Ehrenfest methods. Where we discuss all the relevant issues. In Sec. \ref{sec3}, we discuss our results, figures for the $BEC$ dynamics through a periodic potential embedded on a harmonic confining potential. Moreover, we also discuss the impact of the $PT$-symmetric system over the thermalization of the $BEC$. We observe the impact of the complex part of the potential in the $PT$-symmetric Hamiltonian. Conclusion comes in Sec. \ref{sec4}, with the future suggestions for the related research, and we compare the $BEC$ dynamics with and without the $PT$-symmetric.
 
 \par
\section{Theoretical Models}\label{sec2}
To accurately model an elongated $BEC$, we use a dimensionless quasi-1D  Gross-Pitaevskii equation (GPE) \cite{Pitaevskii03}.
The dimensionless equation can be achieved by taking time $t$ in $\omega_{x}^{-1}$, and scale length $L$=$\sqrt{\hbar/{m\omega_{x}}}$ in terms of harmonic oscillator length along $x$-axis and energy is scaled by $\hbar\omega_{x}$,
 
 \begin{eqnarray}
 \iota\frac{\partial\psi(x,t)}{\partial t}=\left[-\frac{1}{2}\frac{\partial^{2}}{\partial x^{2}}+U(x)+g_{s}|\psi|^{2}\right]\psi(x,t),
 \label{eq1}
 \end{eqnarray}
where $\psi(x,t)$ as a dimensionless  macroscopic wavefunction of the $BEC$, 
$t$ and $x$ stands for time and $1D$-space space coordinate, respectively. Here, we use normalized wavefunction such as, $\int|\psi(x,t)|^{2}dx=1$. The $1D$ interaction strength is given as $g_{s}=2N \omega_{r}a_{s}/(\omega_{x}L)$, where $a_s$ describes the s-wave scattering length \cite{PhysRevA.93.033610}, the number of atoms in $BEC$ is represented by $N$, $\omega_{r}$ stands for  the radial frequency component of the harmonic trap \cite{PhysRevA.93.023606}. To study the thermalization and dipole oscillation for an isolated $BEC$, we purpose the trapping potential $U(x)$,

   \begin{eqnarray}
  U(x)=V(x)+\iota W(x) 
    \label{eq2},
  \end{eqnarray}
\begin{figure}
	\includegraphics[height=8cm,width=9cm]{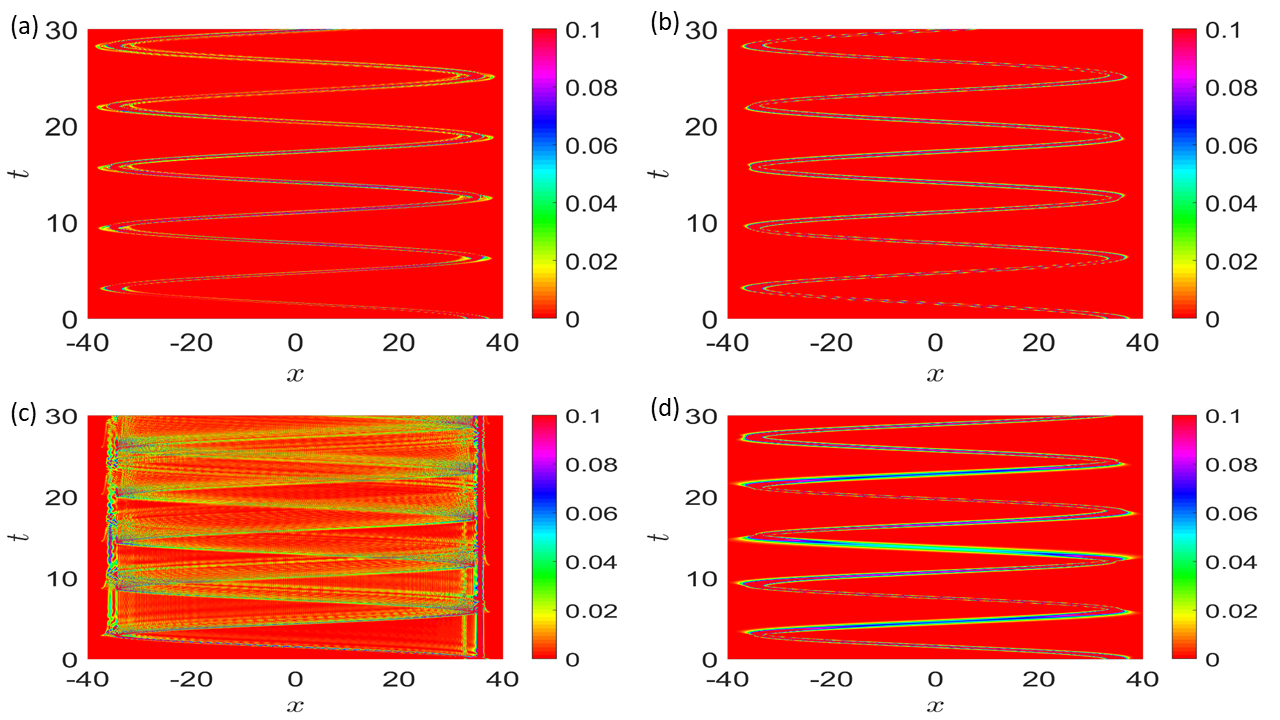}  
	\caption{(Color online) Comparison of  numerical results (left column) and 
		analytical results (right column) of $BEC$ temporal density. The others dimensionless parameters are, interaction strength $g_s=3$, initial mean-position of the BEC defines as $x_0=35$. Here, in upper row the optical periodic potential is $V_0=5$ and in the lower row it consider as $V_0=50$. }
	\label{Fig1}
\end{figure}
 
where the real part of the potential defines as $V(x)=x^2/2+V_0 \cos^2(x)$, here the first term represents the dimensionless harmonic potential confinement, second term models the periodic lattice potential of the system, which serves as an optical lattice of strength $V_0$ and offers   friction to the $BEC$ during its dipole oscillation. The complex part, $W(x)=W_0\sin(x)$ compensate the gain and loss of the $BEC$ atoms, such potentials makes our system a non-Hermitian system. However, this special potential follows the  $PT$-symmetric condition. Here $W(x)<0$ represents the loss of atoms and $W(x)>0$ describes the gain of the $BEC$ atoms.

 \begin{figure}
	\includegraphics[height=8cm,width=9cm]{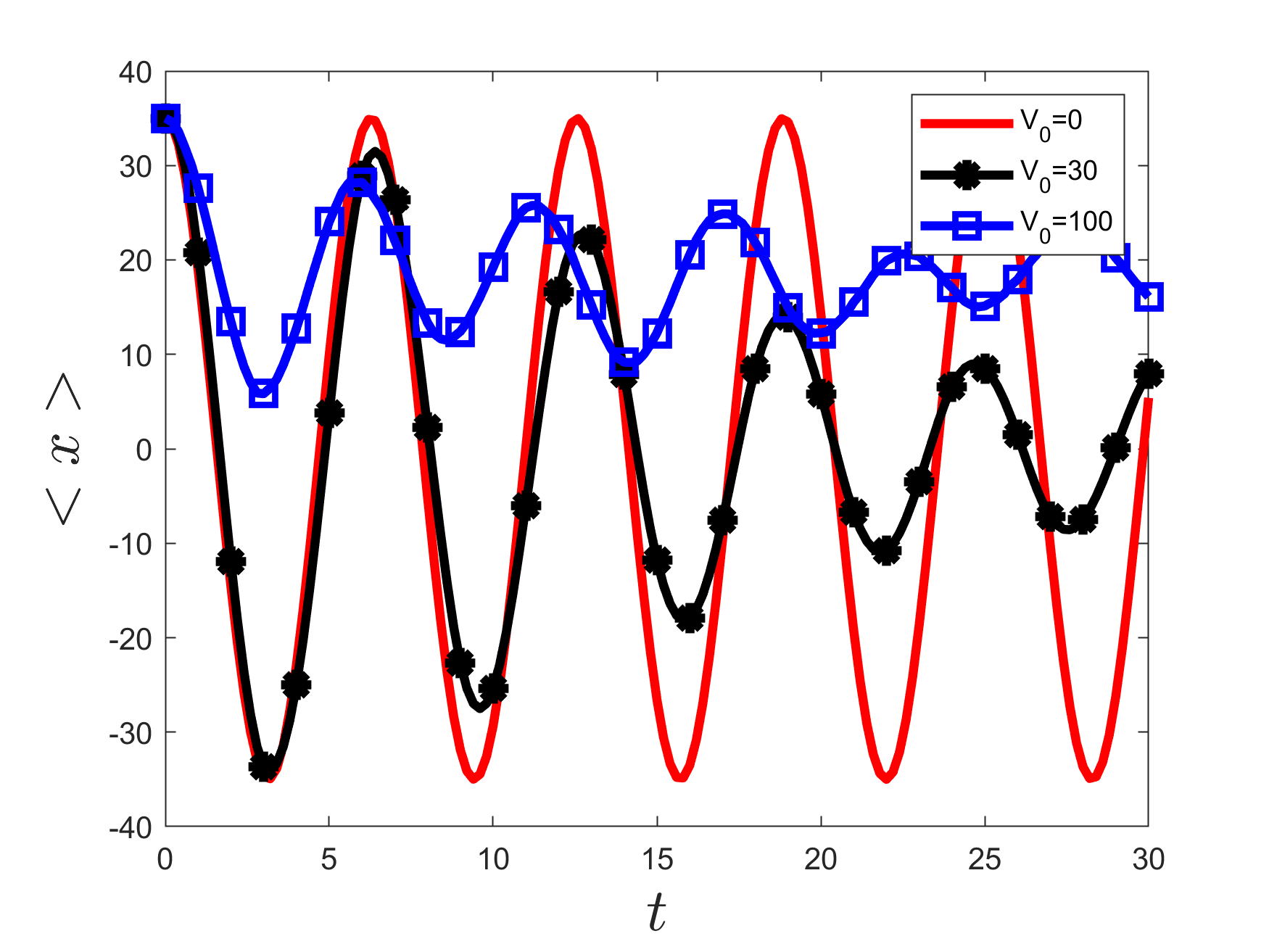} 
	\caption{(Color online) The dimensionless mean position of the  $BEC$ versus the dimensionless time for different periodic potentials $V_0$.   Other dimensionless parameters are $W_0=0$,   $g_s=3$ and the initial mean position of the $BEC$ is $X_0=35$.}
	\label{Fig02}
\end{figure}

\begin{figure*}
	\includegraphics[height=6cm,width=16cm]{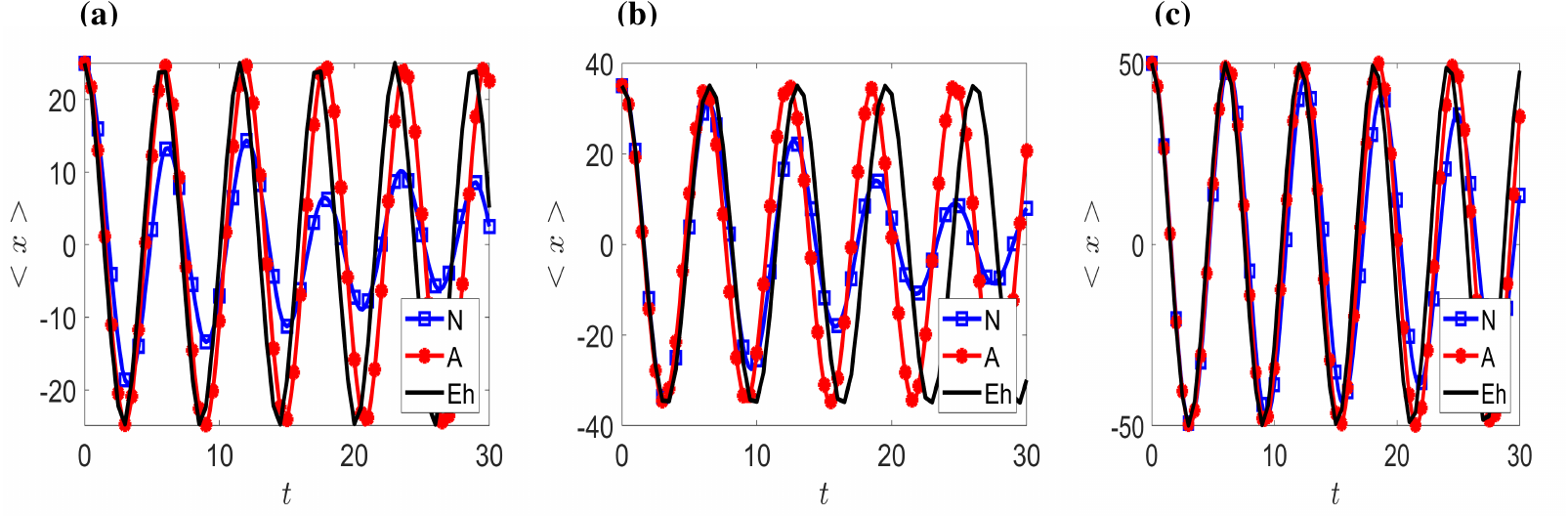}
	\caption{(Color online) The dimensionless mean position of the $BEC$ is plotted against the dimensionless time $t$ for different initial mean position of the $BEC$  $x_0=25$ (a), $x_0=35$ (b), and $x_0=50$ (c).   The mean position is calculated by numerical (blue box), analytical (red circle), and Ehrenfest (black line) methods.  Other dimensionless parameters are, $W_0=0$, $g_s=3$, and the  strength of the periodic potential is $V_0=30$.}
	\label{Fig03}
\end{figure*}

\subsection{Analytical method} \label{sec2-A}
The solution of time dependent $GPE$ by a variational approach \cite{PhysRevA.27.3135,PhysRevA.56.1424,PhysRevLett.77.5320,Borghi_2018} can help to extract the qualitative and quantitative information about the system. 
The variational approach relies on the initial choice of the trial wave function. In our case, we use Gaussian shape wavefunction  with time dependent variables. This approach helps us to find second order ordinary differential equations for the time dependent variables. Which in turn characterize the  dynamics of the $BEC$. Here, we let the initial ansatz as 
\begin{eqnarray}
\psi(x,t)= \frac{1}{\sqrt{a(t)\sqrt{\pi}}} e^{-\frac{(x-x_0(t))^2}{2a(t)^2}+i x \alpha(t)+ix^2\beta(t)},
\label{eq3}
\end{eqnarray}
above ansatz is a Gaussian distribution  centered at $x_0(t)$. Here,  $x$  is defined as the dimensionless space coordinate, $x_0(t)$ describes the dimensionless mean position of the $BEC$, $a(t)$ tells us about the dimensionless width of the $BEC$, $\alpha(t)$ and $\beta(t)$ are the variational parameters. To find all the unknown variational parameters, we let the  Lagrangian density of our system as,
\begin{eqnarray}
\mathcal{L} = &\frac{i}{2}\left(\psi \frac{\partial \psi^*}{\partial t} -
\psi^* \frac{\partial \psi}{\partial t} \right) - \frac{1}{2}|\frac{\partial\psi }{\partial x}|^2 + U(x)|\psi|^2\nonumber \\
&+\frac{g_s}{2} | \psi|^4. \label{eq4}
\end{eqnarray}
Using the above Lagrangian density and the trial wavefunction, we find the effective Lagrangian $L=\int\mathcal{L}dx$  of the quantum mechanical system. We begin by writing the total Lagrangian of the system as a sum of two terms, i.e., $L=L_c+L_{nc}$, where $L_c$ represents the conservative  term  and $L_{nc}$ defines the non-conservative term. The conservative part of the Lagrangian  means that we only consider the real part of the external potential. While the non-conservative term describes the complex part of the external potential. By using the Lagrangian $L$, we determine the complex Ginzburg-Landau equation (CGLE) as \cite{RevModPhys.74.99,PhysRevE.92.022914, Hussain2020}

\begin{eqnarray}
\frac{d}{dt}\Bigg(\frac{ \partial L_c}{\partial \dot{s}}\Bigg)-\frac{ \partial  L_c}{ \partial s}=2Re\Bigg[\int\limits_{-\infty}^\infty i W(x)\psi^\ast \frac{\partial \psi}{\partial s}dx\Bigg],
 \label{eq5}
\end{eqnarray}
where $s$ describe the set of dimensionless  variational parameters such as, $x_0(t)$, $\alpha(t)$, $a(t)$, and $\beta(t)$.
By using equation (4) and equation (5), we determine time-dependent equation for the  dimensionless mean position of the $BEC$, 

\begin{eqnarray}
x_0''(t)+x_0(t)={V_0}\sin(2x_0(t))e^{-a^2(t)}, \label{eq6}
\end{eqnarray}
here, we let $W(x)=0$. While the dynamics of the 
width of the $BEC$  analytically is given by

\begin{equation}
 a''(t)+a(t)=\frac{1}{a^3(t)}+ \frac{g_s}{\sqrt {2\pi} a^2(t)} 
  +2a(t)V_0\cos(2x_0(t))e^{-a^2(t)}.
  \label{eq7}
\end{equation}
For above equations (\ref{eq6}) and (\ref{eq7}) we deliberately avoid to write   the complex part of the potential, i.e., we have only consider the conservative $L_c$ part of the Lagrangian while the "non-conservative" part makes our equations cumbersome, therefore, we avoid to present those lengthy equation here.  
\subsection{Ehrenfest method} \label{sec2-B}
The classical approximation of a quantum system can be realized by the Ehrenfest method \cite{Ehrenfest1927a},
\begin{eqnarray}
 \ddot{\left\langle x \right\rangle }=-\left\langle V^{\prime}(x) \right\rangle, 
\label{eq8}
\end{eqnarray} 
Where $V(x)$ defines the trapping potential for the $BEC$ wave packet and the Ehrenfest theorem leads to the dimensionless mean position of the $BEC$ equation as
\begin{eqnarray}
<\ddot x>=-<x>+2V_0\cos{(<x>)}\sin{(<x>)}, 
\label{eq10}
\end{eqnarray} 
the mean position of the wave-packet is strongly depends on the optical lattice potential $V_0$. 
\subsection{Numerical method}\label{sec2-C} 
To solve numerically the quasi-1D GPE, we use the time-splitting spectral method \cite{BAO2003318} 
 We choose time step as $\bigtriangleup t=0.0001$, and a space step as  $\bigtriangleup x=0.0177$, to discretize the dimensionless quasi-1D GPE Eq.~(\ref{eq1}). 
To give a momentum kick to the $BEC$ wave packet, initially we trap the $BEC$ at potential $V(x)=(x-x_0)^2/2$, where $x_0$ defines the initial mean position of the BEC. Later, we switched off the trapping potential and switch the potential minimum to the new potential $V(x)=x^2/2+V_0 \cos^2(x)$. In this way, $BEC$ experience a kick and starts dipole oscillations in the left over potential.

\begin{figure}
	\includegraphics[height=12cm,width=8cm]{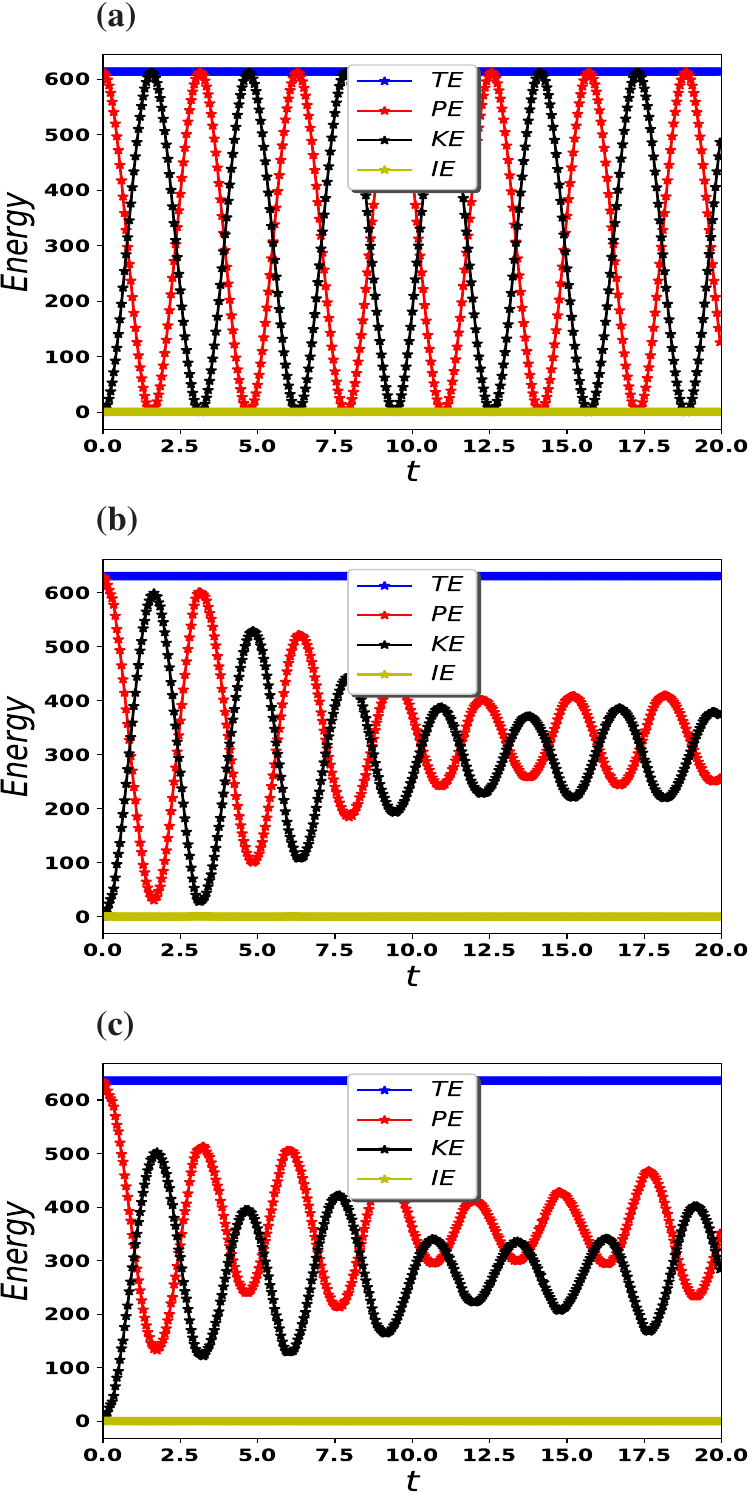}
	\caption{(Color online) Energy evolution of a Bose-Einstein Condensate. Total energy is conserved in the whole process and kinetic energy and potential energy associated with harmonic trapping are evenly distributed at equilibrium. The parameters used are (a)$V_0=0$, (b)$V_0=30$, and (c)$V_0=40$.}
	\label{Fig04}
\end{figure}

\section{$BEC$ dynamics}\label{sec3}
To study the dynamics of a $BEC$ in this closed environment,  we compare the  analytical, Ehrenfest, and numerical methods discussed previously in  Sec. \ref{sec2} (A-C).
First of all, the qualitative comparison of analytical and numerical results are presented in Fig.~\ref{Fig1} in the form of a temporal density plot of the $BEC$ in the absence of $PT$-symmetric potential. $BEC$ dynamics applied under a $PT$-symmetric potential environment is presented in Fig.~\ref{Fig05}. We discuss both cases separately in the following two subsections. 

\subsection{Without $PT$-symmetric Potential}\label{sec3-A}
In this subsection, we compare and discuss the detailed result of the $BEC$ dynamics without $PT$-symmetric potential, i.e., $W_0=0$, using  analytical, Ehrenfest, and numerical methods. 
\subsubsection{Density dynamics of the  $BEC$} \label{sec3-B}
The temporal density graph for the $BEC$ is obtained numerically and analytically  as  shown in Fig.~\ref{Fig1}.     
For small values of the periodic potential strength, $V_0=5$, both analytical  and numerical results are in agreement with each other as shown in Fig.~\ref{Fig1}(a-b).
However, we observe that for higher values of the periodic potential strength, e.g., $V_0=50$, as presented  in Fig.~\ref{Fig1}(c-d) both plots differ from each other. We note that the results obtained by the analytical method, Fig.~\ref{Fig1}(d), fails to reflect the physics of the dynamics of the $BEC$. Particularly, it shows no impact of scattering of the $BEC$ due to the friction offered by the lattice potential. While the results obtained by the numerical method, Fig.~\ref{Fig1}(c), reveal  the impact of the scattering from the peaks of the lattice potential and the dissipation. It reveals that quasi-particles are generated at the top of the BEC, which leads to thermalization of the BEC. 
 So from now, for this section, we discuss only numerical results, because they present the real picture of the dynamics of the $BEC$.
  
For further investigation, we plot the mean position of the $BEC$, by using the  numerical method as shown in Fig.~\ref{Fig02}. 
We observe that for periodic potential strength $V_0=0$  the $BEC$ starts dipole oscillation without any dissipation in the closed environment as shown in Fig.~\ref{Fig02}.    We note that for $V_0=30$, in Fig.~\ref{Fig02},
the numerical results shows that the mean position of the $BEC$ starts localizing at the global minima of the harmonic potential trap as presented in Fig.~\ref{Fig02}. For such a lattice potential the mean position dipole oscillation has a smaller amplitude.  
\textbf{ For a larger value of the periodic potential, $V_0=100$ small $BEC$ dipole oscillation can be seen as depicted in Fig.~\ref{Fig02}. }
Initially, BEC starts moving towards the global minima of the external potential but on the way, it loses its energy and turns back, even without reaching the global minima of the potential and after some time it gets localized at   local minima as shown in Fig.~\ref{Fig02}. This localization other than the global minima of the harmonic potential is due to the loss of the energy of the $BEC$ by the periodic potential embedded on the harmonic potential. It is also quite surprising that the tunneling of the $BEC$ is also suppressed in this special scenario, however, eventually, the BEC will localize to global minima but after a long time.  

\subsubsection{The mean position vs initial energy of the $BEC$} \label{sec3-C}
The qualitative comparison of the dimensionless mean position of the BEC is plotted in  Fig.~\ref{Fig03}, for three different methods, numerical (N), analytical (A), and Ehrenfest (Eh). We compare the impact of the initial potential energy ($PE)$ of the $BEC$ on its dipole oscillations  
 without any $PT$-symmetric environment.  
Under such condition, it is evident from the Fig.~\ref{Fig03} that the initial potential energy of the $BEC$ depends upon the choice of the initial mean position of the   $BEC$, $``x_0''$. We know that  the dimensionless potential energy is given by, $PE\propto x_0^2$.  
The initial energy of the $BEC$ shows a considerable influence on the dipole oscillations as plotted in Fig.~\ref{Fig03}. 
In Fig.~\ref{Fig03}, we see the influence of initial energy on the dipole motion of the $BEC$  
 for $x_0=25$, the numerical study shows that the dissipation of the $BEC$ results in an earlier localization of the mean position of the  $BEC$.  
On the other hand, for higher values of the magnitude of $PE$, say $x_0=50$, 
the $BEC$ just experience the global harmonic potential and hence a to-and-fro motion results. While the initial high $PE$ compensates the periodic frictional potential. 
We note that for low initial potential energy, the $BEC$ localized earlier 
as presented in Fig.~\ref{Fig03}. We also observe that the Ehrenfest and analytical methods could not capture the physics of the dimensionless mean-position of the $BEC$. As we do not find the dependence of the mean-position on the initial PE of the system, which is not physical. However, from the  numerical calculation, we conclude that the initial high $PE$ of the $BEC$  results in a delay in the localization of the mean position of the $BEC$. We observe that the $BEC$ with high $PE$ maintains longer dipole oscillation for a longer time. Therefore, it is appropriate for experimentalists  to consider this point while localizing the $BEC$.  
i.e., they must not put  
the $BEC$ far away from the global minima, as it could lead to decoherence in the experiment, which could destroy the $BEC$.
In a classical harmonic system, the total energy of the system is dissipative, due to the environment interaction, however, in our special case the system is isolated therefore the total energy is conserved as shown in Fig.~\ref{Fig04}. It means that periodic potential does not store energy during the thermalization process, however, periodic-potential distribute the energy, few other writers have also studied this phenomenon in disorder potentials \cite{Hsueh2020}. As someone can see from   Fig.~\ref{Fig04} that for higher values of $V_0$ the kinetic energy (KE) and potential energy (PE) are  oscillating, however, as time  passes oscillation becomes smaller and smaller which leads to equilibrium.  As a matter of fact, this energy distribution indicates another kind of non-equilibrium to an equilibrium state. 
 
Since our system is in an isolated environment. So the localization is quite surprising in such a system. However, someone can answer this localization is due to the thermalization of the $BEC$. As the $BEC$ is started to move from its initial position, it experiences friction in the system due to periodic potential. That periodic resistance generated quasi-particles at the top of the $BEC$, which can be seen in Fig.~\ref{Fig1}(c). This is not new as many studies already pointed out such quasi-particles due to the collision of the $BEC$  wave-packet with external potentials \cite{PhysRevA.82.033603}.

 \subsection{With $PT$-Symmetric Environment} \label{sec3-D}
 
In this sub-section, we compare analytical and numerical  $BEC$ dynamics by observing the temporal density and the mean position of the $BEC$ under a $PT$-symmetric environment. The amount of $PT$-symmetry or the imaginary part of the potential controlled by the parameter, $W_0$. 
 \subsubsection{The temporal density graph}
\begin{figure}
\includegraphics[height=8cm,width=9cm]{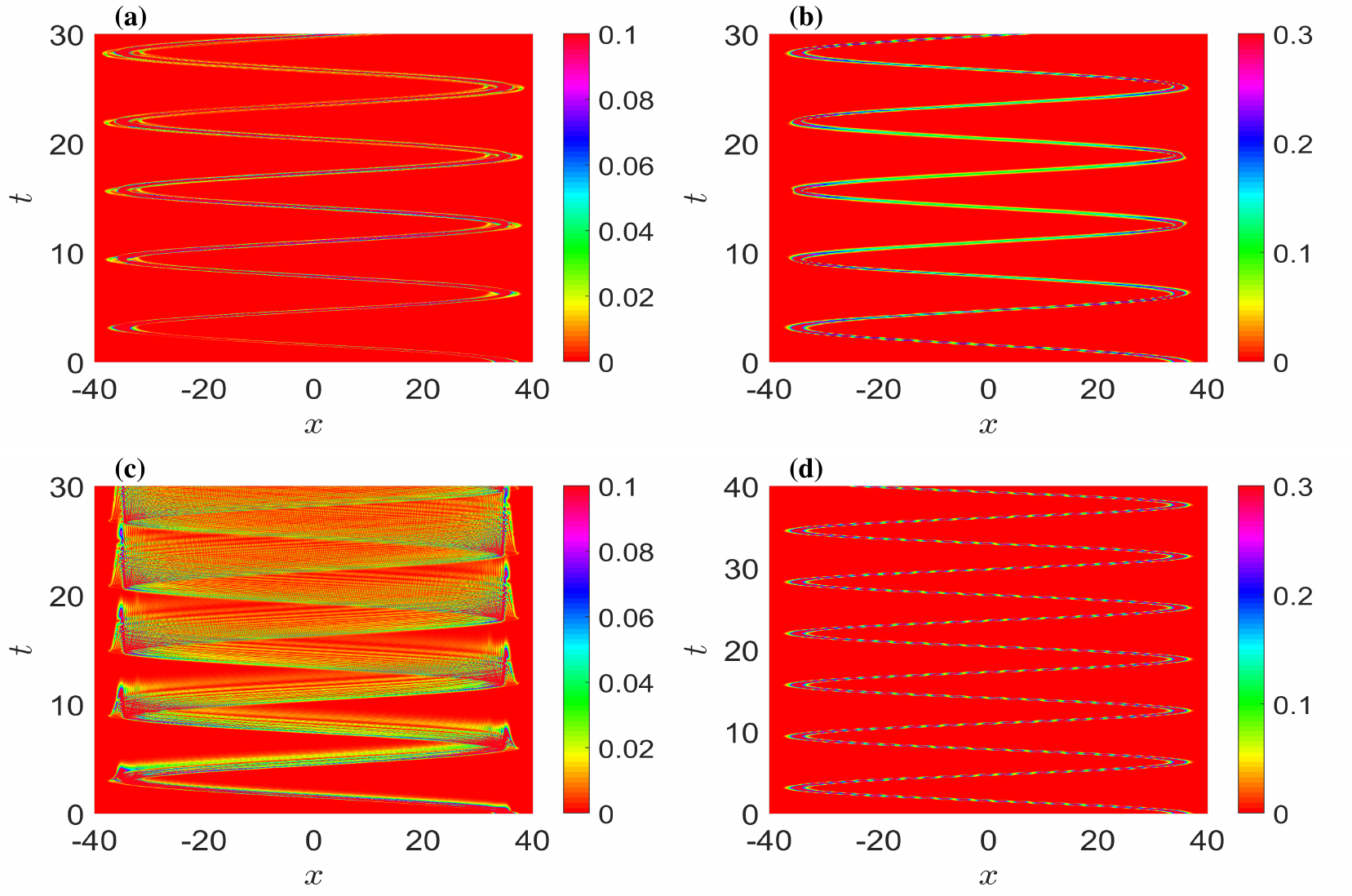} 
\caption{(Color online) Comparison of numerical (left column) and  analytical (right column)  results of the $BEC$ temporal density with a $PT$-symmetric potential. The dimensionless parameters defines as the interaction strength $g_s=3$, initial mean position of the $BEC$ $x_0=35$. The periodic potential strength for upper row is $V_0=5$  and the strength of the imaginary part of potential is $W_0=0.2$, while for the lower row the periodic potential strength is $V_0=30$ and the strength of the imaginary part of the potential is $W_0=0.2$.}
\label{Fig05}
\end{figure}

The temporal density graph in Fig.~\ref{Fig05} is obtained by analytical (right column) and numerical simulation (left column). Here, we note that for a small amount of $PT$-symmetry, $W_0=0.2$ and for a small amount of the strength of the external periodic potential, $V_0=5$, the analytically and numerically obtained results for the dynamics of the $BEC$ are in good agreement. 
While there is disagreement in the results for higher values of the periodic potential strength, like $V_0=30$. For larger values of $V_0$ once again, we observe that the numerical results are in agreement with the physics of the dissipative dynamics and analytical methods are no longer valid. 

\begin{figure}	
	\includegraphics[height=6cm,width=9cm]{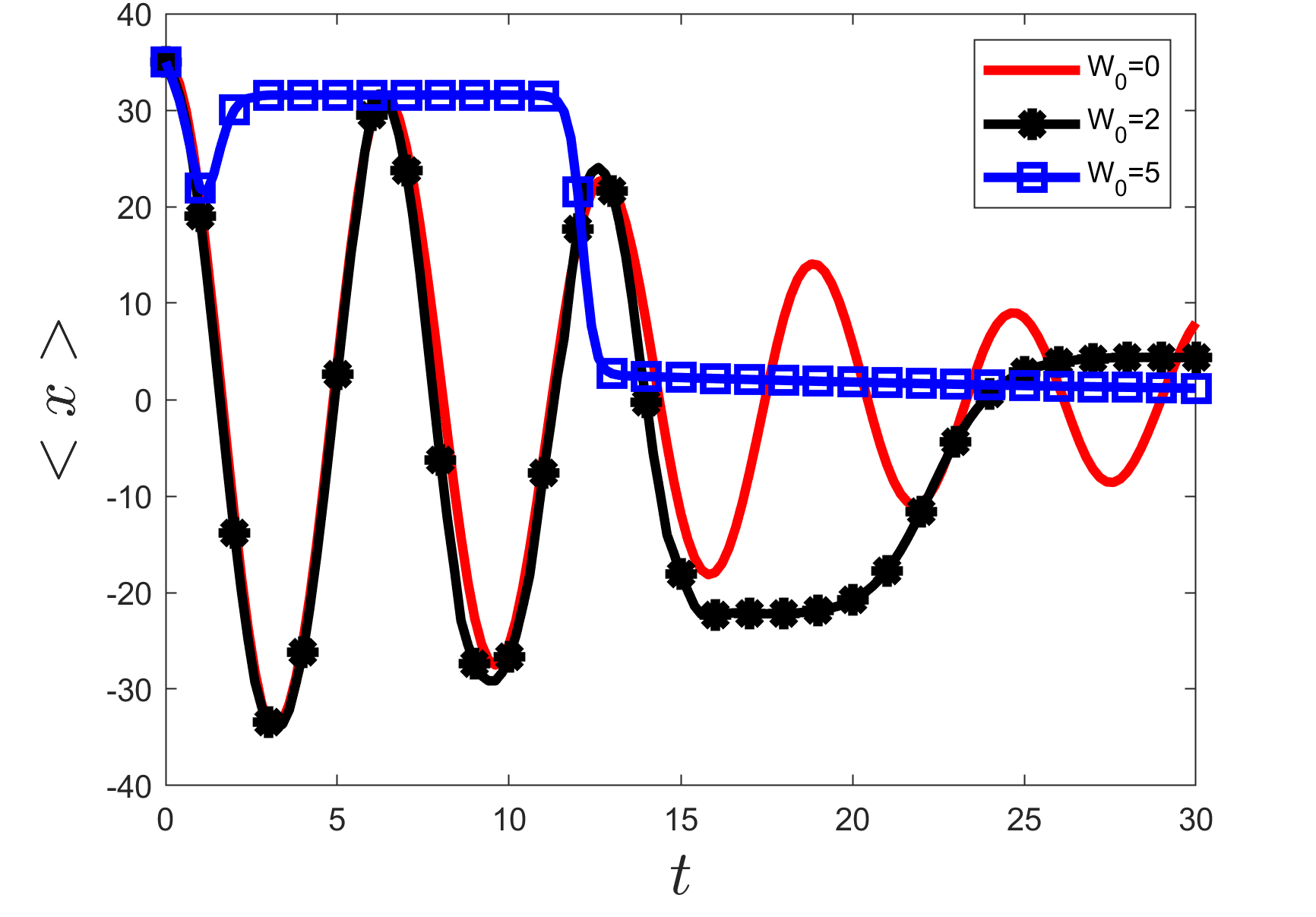}  
	\caption{(Color online) The impact of the amount of  $PT$-symmetric potential on mean position of the $BEC$.  The $BEC$ initially located at $x_0=35$. Other dimensionless parameters are, $g_s=3$, $V_0=30$.}
	\label{Fig06}
\end{figure}
   
\subsubsection{Temporal mean position}

In this subsection, we discuss the dynamics of the dimensionless mean position of the $BEC$, which is calculated by numerical method.
%

We observe that for small $PT$-symmetric potential the $BEC$ shows the famous dipole oscillation, with a gradual decrease in dipole oscillation's amplitude as shown in Fig.~\ref{Fig06}. Which describes the localization of the $BEC$ as already discussed in Sec. \ref{sec2}. As we increase, the $PT$-symmetric potential "$W_0=2$" we observe that the mean of the $BEC$ starts dipole oscillation but around "$t=15$" we note that the dipole oscillation rapidly seized and the $BEC$ localized at some local minima. However, as time passes the $BEC$ mean position jumps to the global minima as predicted in Fig.~\ref{Fig06}. We realize  that this jump is quite natural as in $BEC$ particles are continuously ejected from the adjacent wells simultaneously. Additionally, there is a continuous compulsion for the $BEC$ to move towards the global minima.
As we raise the strength of the $PT$-symmetric to "$W_0=5$" we inspect that the mean point of the $BEC$  is localized around "$x=30$, in a relatively short time as shown in Fig.~\ref{Fig06}.  And it stays in this local minima for a relatively long time. Later, the mean position of the $BEC$ switches towards the global minima. It is quite strange that we could not see the localization of the $BEC$ in between the "$x=30$" and "$x=0$". This is quite interesting as it leads to a kind of digital switching. So with this, we conclude that our proposed model can be used to study the localization of the $BEC$,  additionally, it can also be used for   discrete switching.
 \section{conclusion}\label{sec4}
In this paper, we compare the analytical, numerical, and Ehrenfest methods to study the $BEC$  dynamics in a dissipative environment. Along with this, we also study the impact of the presence of the $PT$-symmetric on the dissipative dynamics of  the $BEC$. The dissipative environment is created by adding a periodic potential over a harmonic potential. The dissipation is controlled by the strength of the periodic potential height. 
We conclude that the analytical and Ehrenfest methods have limitations for  larger values of the periodic potential strength, $V_0$ . For larger values of the periodic potential strength, the numerical methods remain  valid.
The presence of a periodic  $PT$-symmetric environment influences the dynamics of the $BEC$ in such a way that it can control the localization of the $BEC$ at the desire  location. By controlling the amount of $PT$-symmertric strength and the strength of the periodic potential parameter we can localize the $BEC$ to a desirable location for a desirable time. \textbf{As a future prospective, someone can extend this work for spin-orbit coupled BEC's \cite{PhysRevLett.108.080406} and for dipolar condensate \cite{PhysRevA.84.041604,PhysRevA.88.013624}}.

\section{Acknowledgment}
Jameel Hussain gratefully acknowledges support from the COMSATS University Islamabad for providing him a workspace. 
 
\bibliographystyle{apsrev4-1}

%

\end{document}